# The reaction-diffusion approach to morphogenesis

Rui Dilão
Non-Linear Dynamics Group
IST, Department of Physics
Av. Rovisco Pais, 1049-001 Lisbon, Portugal
rui@sd.ist.utl.pt

*Abstract*—
**Morphogenesis is the ensemble of processes that determines form, shape and patterns in organisms. Based on a reaction-diffusion theoretical setting and some prototype reaction schemes, we make a review of the models and experiments that support possible mechanisms of morphogenesis. We present specific case studies from chemistry (Belousov-Zhabotinsky reaction) and biology (formation of wing eyespots patterns in butterflies). We show the importance of conservation laws in the establishment of patterning in biological systems, and their relevance to explain phenotypic plasticity in living organisms. Mass conservation introduces a memory effect in biological development and phenotypic plasticity in patterns of living organisms can be explained by differences on the initial conditions occurring during development.**

I. INTRODUCTION

Proteins are folded arrays of amino-acids, coded by sequences of base pairs in DNA strands or genes. Proteins are present in all living organisms and are involved in almost all biochemical processes that occur in the cell. They can be simple functional agents with a specific action and participate in the control of biological processes. From the biochemical point of view, each species is characterized by its ensemble of genes or genome.

The spatial form and shape of pluricellular organisms is gradually acquired during its development, which begins in the zygote. After the first stages of replication of the zygote, a sequence of mechanical movements and shape transformations succeed in time, leading to the final form of the organism, [GIL 97]. In some organisms as insects, the shape of the larvae is preceded by a specific pattern of distribution of proteins appearing at an early multinucleated phase of the embryo (syncytium), [NUS 96]. In other organisms as in the haploid amoeba *Dictyostelium discoideum*, the form of the social aggregates or fruiting bodies (myxamoeba) results from a chemotaxis phenomena mediated by cAMP (cyclic adenosine monophosphate), [TOM 81].

In all these cases, the form and shape of organisms is not coded in the genome. Genes encode proteins, and it is difficult to conceive that specific patterns as well as final forms and shapes of an organism could be in direct correspondence with specific proteins. Therefore, morphogenesis, the science of patterns, form and shape, must be of epigenetic nature. On the other hand, the gradual establishment of form and shape in organisms is a dynamic process, dependent on the internal processes occurring in the embryo and also on external factors as temperature, pressure, *etc.*.

There are essentially two main theoretical approaches in the study of morphogenesis. One approach relies on the study of possible mechanical deformations of forms and shapes in three-space. This approach has been pioneered by René Thom [THO 72] and consists in the study of all possible shapes (manifolds) in the three-dimensional Euclidean space and their transformations (bifurcations and catastrophes). The second approach has been proposed by Alain Turing [TUR 52] and links chemical transformations with the flow emerged from the differences of concentration of specific substances between adjacent cells. Both theoretical points of views predict specific effects that are observed in some laboratory experiments and organisms, and are useful to test functional models and mechanisms of development.

An important issue associated with the Turing approach to morphogenesis is the microscopic interpretation of the flows of substances in solution in a media. The concentration flow described by the first Fick´s law can be understood as the result of the Brownian motion of the molecules in solution, imposing a concentration gradient in the direction of lower concentrations. This simple mechanism, described by the diffusion equation, enables the construction of simple mathematical models based on the independence of the motion of individual molecules in solution and on the random collision that occur at the molecular scale. As a consequence, forms, shapes and patterns are emerging phenomena resulting from the simple inertial laws of motion at the microscopic scale.

It was Turing [TUR 52] who suggested for the first time that, in spatially extended media or biological tissues, the diffusive coupling between reacting substances can generate stable patterns (Turing patterns). These patterns are characterized by a non-homogeneous and steady spatial distribution of the concentration of some substance. In extended media, the fluctuations in the concentration of a substance in solution are eliminated by the intrinsic random movement or Brownian motion of the molecules in the media. The Brownian motion homogenises the local concentrations, preventing the emergence of steady gradients. However, for several reacting substances in a media, the situation can be completely different. For example, two colliding molecules with binding affinity can lead to a third molecule, and the local concentration of the third molecule increases. The analysis of Turing went further, and he suggested that, in multicomponent systems of reaction-diffusion equations, we could also have travelling wave type phenomena. Turing called the diffusion driven reacting substances morphogens or evocators, a term used by Waddington to explain the morphological transformations induced by specific substances in the embryo of the chick, [WAD 40]

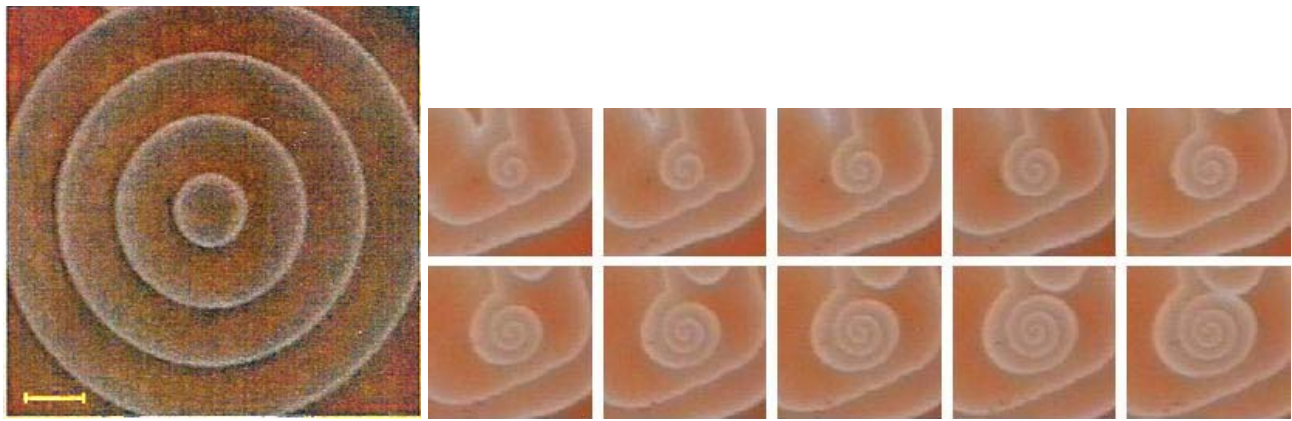

Fig. 1 Wave fronts and spiral waves observed in the Belousov-Zhabotinsky reaction, from [SAI 99]. The wave fronts or target patterns appear spontaneously in the Belousov-Zhabotinsky reaction in extended media, around an "organizing" centre or pacemaker. It is believed that the "organizing" centre can be an impurity in the media or due to a strong local fluctuation in the concentration of one of the substances in solution.

The oscillatory behaviour predicted by Turing has been observed experimentally in the Belousov-Zhabotinsky reaction in extended media [ZAI 70]. This auto-catalytic reaction has been invented as an example of an oscillatory chemical reaction, [ZHA 91]. In homogeneous media and in an open flow regime, the main characteristic of this reaction is the induction of periodic oscillations in the concentrations of the intermediates. In extended media, the local fluctuations in the concentrations of the reacting substances develop wave type fronts that propagate along the media. Manipulating experimentally these wave fronts, it possible to obtain experimentally growing spiral concentration waves, one of the most abundant patterns observed in nature, [ZAI 71], Fig. 1.

The patterns observed in the Belousov-Zhabotinsky reaction, concentric or target waves and spiral waves, are similar to the ones that appear during the processes of aggregation of the haploid amoeba *Dictyostelium discoideum*, [TOM 81] and [GIL 97]. Reaction-diffusion systems can still originate stable or Turing patterns, and the experimental observation of Turing patterns in a chemical experiment is recent, [CAS 90].

The aim of this paper is to present the foundations leading to models of pattern formation based on the Turing approach. In order to maintain the consistency and completeness of this paper, in section II, we discuss the mass action law and the derived reaction-diffusion equations. We also discuss the conditions leading to the existence of travelling wave solutions, spiral wave solutions and Turing patterns solutions of reaction-diffusion equations. We describe briefly the optics of non-linear reaction-diffusion waves and the phenomena of back refraction appearing in reaction-diffusion equations, [SAI 98]. We compare numerical results obtained with prototype models with some qualitative aspects of pattern formation observed in experiments.

In section III, we present a model for the morphogenesis of wing eyespots patterns in butterflies, [DIL 04a]. In this model, diffusion has the role of propagating information along the extended media, and patterns are a consequence of the conservation local laws associated to non-diffusing substances. This model is an extension to the Turing original approach to morphogenesis.

The consequences of the models presented along this paper and the comparison between model results with experiments and observations are discussed in the final section IV. Independently of the accuracy and the dynamic details of the models described here, all the experimental systems discussed here have the signature of the reaction-diffusion type mechanisms. In particular, back refraction and annihilation of colliding wave fronts, simulated by reaction-diffusion systems are present in the Belousov-Zhabotinski reaction, in the butterfly eyespot patterns and in the patterns of aggregates of *Dictyostelium discoideum*.

II. PATTERN FORMATION WITH REACTION-DIFFUSION EQUATIONS

*A. The mass action law and the Turing hypothesis*

In solutions, atoms and molecules collide and may bind if they have the necessary chemical affinity, determined by the nature of the chemical bound of the colliding molecules. Assuming that we have a total number of $m$ chemical substances, and $n$ chemical reactions, the reactions occurring in the media are represented by the $n$ collision diagrams,

$$\nu_{i1} A_1 + ....... + \nu_{im} A_m \xrightarrow{k_i} \mu_{i1} A_1 + ....... + \mu_{im} A_m \quad (i=1,...,n) \tag{1}$$

where $A_j$ represents the atom or molecule of substance number $j$, and the $k_i$'s are rate constants. In the following, we use the same letter to represent the atom or molecule of a substance and its concentration in the media. The numbers $\nu_{ij}$ and $\mu_{ij}$ are the stoichiometric coefficients (integers). The order of each reaction in (1) is $r_i = \sum_{j=1}^{m} \nu_{ij}$. These diagrams represent a collision between different molecules. For example, if we have three molecules, and molecule $A_1$ collides with molecule $A_2$, forming molecule $A_3$, the corresponding reaction diagram is,

$$1A_1 + 1A_2 + 0A_3 \xrightarrow{k} 0A_1 + 0A_2 + 1A_3$$

or, in the simplified form,

$$A_1 + A_2 \xrightarrow{k} A_3$$

and $k$ is the rate of formation of $A_3$. If, in the collision diagram number $i$, $r_i = 2$, we have a binary collision or a reaction of order $2$. If, $r_i > 2$, we have a higher order collision. Collisions involving three or more molecules have a very low probability of occurrence, and in chemical kinetics they are not in general considered.

To derive the equations for the time evolution of the concentrations of the chemical substances in a given ensemble of chemical transformations of type (1), it is assumed that:
  i) Substances are homogeneous on the media and have low density.
  ii) All reactions occur at constant volume and temperature.
  iii) The individual motion of the molecules in the media is independent from the other molecules, behaving as Brownian particles, and their collision frequency is proportional to the local concentration.

Under these conditions, the time evolution of the concentrations of all the chemical substances represented by the generic diagram (1) is described by the set of $m$ ordinary differential equations,

$$\frac{dA_j}{dt} = \sum_{i=1}^{n} k_i (\mu_{ij} - \nu_{ij}) A_1^{\nu_{i1}} ... A_m^{\nu_{im}} \quad (j = 1,...,m) \tag{2}$$

where $A_j$ represents the concentration of chemical substance number $j$. The system of equations (2), calculated from the diagrams (1), is called the law of mass action, ([HAK 83], [VOL 94]).

However equations (2) are not necessarily independent. Writing equations (2) in the form,

$$\frac{d}{dt}\begin{pmatrix} A_1 \\ \vdots \\ A_m \end{pmatrix} = \begin{pmatrix} \mu_{11} - \nu_{11} & \cdots & \mu_{n1} - \nu_{n1} \\ \vdots & & \vdots \\ \mu_{1m} - \nu_{1m} & \cdots & \mu_{nm} - \nu_{nm} \end{pmatrix} \begin{pmatrix} k_1 A_1^{\nu_{11}} ... A_m^{\nu_{1m}} \\ \vdots \\ k_n A_1^{\nu_{n1}} ... A_m^{\nu_{nm}} \end{pmatrix} := \Gamma \begin{pmatrix} \omega_1 \\ \vdots \\ \omega_n \end{pmatrix} = \Gamma \omega \tag{3}$$

the $n \times m$ matrix $\Gamma$ has rank $r$, and $r \leq \min\{n, m\}$. If $r = m$, then the equations in (3) are linearly independent. If $r < m$, equations in (3) are linearly dependent. To analyze this case, let $V_j^T = (\mu_{1j} - \nu_{1j},...,\mu_{nj} - \nu_{nj})$, with $j = 1,...,m$, denote the row vectors of $\Gamma$. Then, as $r < m$, there exists constants $\alpha_{jk}$, with $j = 1,...,m$ and $k = 1,...,m-r$, such that $\sum_{j=1}^{m} \alpha_{jk} V_j = 0$. As each equation in (3) can be written as, $\frac{dA_j}{dt} = V_j . \omega$, we have, $\sum_{j=1}^{m} \alpha_{jk} \frac{dA_j}{dt} = \sum_{j=1}^{m} \alpha_{jk} V_j . \omega = 0$, implying that,

$$\sum_{j=1}^{m} \alpha_{jk} A_j = \text{constant}_k \quad k = 1,...,m-r \tag{4}$$

Relations (4) define $m - r$ conservation laws,

$$\sum_{j=1}^{m}\alpha_{jk}A_j(t)=\sum_{j=1}^{m}\alpha_{jk}A_j(0) \quad k=1,\ldots,m-r \tag{5}$$

Turing implicitly assumed that all the reactions occurring in a media obey the mass action law, and as in biological tissues concentrations are away from spatial homogeneity, diffusion may play an important role in the establishment of growing fluctuation in the local concentration of some chemical species. Some substances diffuse along tissues, and others stay localized. The diffusing substances were called morphogens, form producers, or evocators.

Taking the Turing hypothesis to all its full consequences, the spatial form and shape originated by a given chemical mechanism is described by the set of reaction-diffusion equations,

$$\frac{d}{dt}\begin{pmatrix}A_1\\ \vdots \\ A_m\end{pmatrix}=\begin{pmatrix}\mu_{11}-\nu_{11} & \cdots & \mu_{n1}-\nu_{n1}\\ \vdots & & \vdots\\ \mu_{1m}-\nu_{1m} & \cdots & \mu_{nm}-\nu_{nm}\end{pmatrix}\begin{pmatrix}k_1 A_1^{\nu_{11}}\ldots A_m^{\nu_{1m}}\\ \vdots\\ k_n A_1^{\nu_{n1}}\ldots A_m^{\nu_{nm}}\end{pmatrix}+\begin{pmatrix}D_1\Delta A_1\\ \vdots\\ D_n\Delta A_n\end{pmatrix} \tag{6}$$

where the $D_i$'s are diffusion coefficients associated to the morphogens or form producers, and $\Delta=\left(\frac{\partial^2}{\partial x_1^2}+\ldots+\frac{\partial^2}{\partial x_k^2}\right)$ is the $k$-dimensional Laplace operator. In order to obtain a realistic model for a specific mechanism of morphogenesis, we must identify the morphogenic substances with $D_i>0$, and the non-diffusing substances ($D_j=0$). This implies a very precise knowledge of the developmental process under analysis. However, Turing followed a more schematic approach.

Turing discovered that the coupling through diffusion of a system of two non-linear ordinary differential equations with a stable fixed point, can lead to the instability in the eigenmodes of the associated linearized reaction-diffusion equation. He built a one-dimensional spatial system consisting in a finite number of connected cells arranged in a ring. In each cell, chemical substances evolve in time according to the law of mass action (2). The flow between adjacent cells is proportional to the difference of local concentrations, and the spatial extended system is described by a reaction-diffusion system of equations.

To be more specific, Turing considered the system of reaction-diffusion equations,

$$\begin{aligned}\frac{\partial \varphi_1}{\partial t}&=f(\varphi_1,\varphi_2)+D_1\Delta\varphi_1\\ \frac{\partial \varphi_2}{\partial t}&=g(\varphi_1,\varphi_2)+D_2\Delta\varphi_2\end{aligned} \tag{7}$$

where $D_1$ and $D_2$ are the diffusion coefficients of $\varphi_1$ and $\varphi_2$, respectively, and the spatial domain is the $k$-dimensional cube of side length $S$. In general, in the literature of reaction-diffusion systems, as well as in the Turing paper, it is implicitly assumed that the local equations in (7) are independent and the phase space variables and parameters are independent.

If the local system associated to (7) has a fixed point at $(\varphi_1,\varphi_2)=(0,0)$, linearizing (7) around $(\varphi_1,\varphi_2)=(0,0)$, we obtain,

$$\frac{d}{dt}\begin{pmatrix}\varphi_1\\ \varphi_2\end{pmatrix}=\begin{pmatrix}a_{11} & a_{12}\\ a_{21} & a_{22}\end{pmatrix}\begin{pmatrix}\varphi_1\\ \varphi_2\end{pmatrix}+\begin{pmatrix}D_1\Delta\varphi_1\\ D_2\Delta\varphi_2\end{pmatrix} \tag{8}$$

For initial data obeying Neumann boundary conditions (zero flux), the solutions of (8) have the general form,

$$\varphi_1(x_1,...,x_k,t) = \sum_{n_1,...,n_k \geq 0} c_{n_1,...,n_k}(t) \cos(\frac{2\pi n_1}{S} x_1)...\cos(\frac{2\pi n_k}{S} x_k)$$

$$\varphi_2(x_1,...,x_k,t) = \sum_{n_1,...,n_k \geq 0} d_{n_1,...,n_k}(t) \cos(\frac{2\pi n_1}{S} x_1)...\cos(\frac{2\pi n_k}{S} x_k)$$

(9)

where $c_{n_1,...,n_k}(t)$ and $d_{n_1,...,n_k}(t)$ are the Fourier coefficients of the solution of (8). The terms under the sum in (9) are the eigenmode solutions of (8), and are indexed by a $k$-tuple of non-negative integers. Introducing (9) into (8), we obtain the infinite system of ordinary differential equations,

$$\frac{d}{dt}\begin{pmatrix} c_{n_1,...,n_k} \\ d_{n_1,...,n_k} \end{pmatrix} = \begin{pmatrix} a_{11} - 4D_1 \frac{\pi^2}{S^2}(n_1^2 +...+ n_k^2) & a_{12} \\ a_{21} & a_{22} - 4D_2 \frac{\pi^2}{S^2}(n_1^2 +...+ n_k^2) \end{pmatrix} \begin{pmatrix} c_{n_1,...,n_k} \\ d_{n_1,...,n_k} \end{pmatrix}$$

$$:= J_{n_1,...,n_k} \begin{pmatrix} c_{n_1,...,n_k} \\ d_{n_1,...,n_k} \end{pmatrix}$$

(10)

with $(n_1,...,n_k) \geq (0,...,0)$. For each non-negative $k$-tuple of integers $(n_1,...,n_k)$, the stability of the eigenmode solutions of (10) is determined by the eigenvalues of the matrix $J_{n_1,...,n_k}$. Writing the eigenvalues of $J_{n_1,...,n_k}$ as a function of the trace and determinant, we obtain,

$$\lambda_{n_1,...,n_k}^{+-} = \frac{1}{2}\left(TrJ_{n_1,...,n_k} \pm \sqrt{(TrJ_{n_1,...,n_k})^2 - 4DetJ_{n_1,...,n_k}}\right)$$

(11)

By (10), for every $(n_1,...,n_k) \geq (0,...,0)$, the real and imaginary parts of (11) are bounded from above, and we can define the number,

$$\Lambda = \max\{\text{Re}(\lambda_{n_1,...,n_k}^{+-}) : (n_1,...,n_k) \geq (0,...,0)\}$$

(12)

The number $\Lambda$ is the upper bound of the spectral abscissas of the set of matrices $\{J_{n_1,...,n_k} : (n_1,...,n_k) \geq (0,...,0)\}$.

This linear analysis lead Turing to find that, in the case of two or more diffusive and reacting substances, a stable state of the local system (8) could be destabilized by the diffusion terms, inducing a symmetry breaking in the global behavior of the solutions of the non-linear system (7). This effect is called today Turing or diffusion-driven instability, [CRO 93]. Depending on the type and magnitude of the unstable eigenmodes of the linearized reaction-diffusion system, different asymptotic states of the non-linear system could eventually be reached. Turing implicitly conjectured that if the eigenvalue with dominant real part of all the unstable eigenmodes of the linearized system is real, an asymptotic time independent solution of the non-linear equation could eventually be reached. On the other hand, if the eigenvalue with dominant real part of all the unstable eigenmodes is complex, the solution could evolve into a time periodic spatial function or wave. In the first case, we are in the presence of a Turing instability and, in the second case, we have an oscillatory instability. The Turing instability can then be defined in the following way:

**Definition** (Turing instability): Near the fixed point $(\varphi_1, \varphi_2) = (0,0)$, the reaction-diffusion system (7) has a Turing or a diffusion-driven instability of order $(n_1,...,n_k) \geq (0,...,0)$, if the upper bound of the spectral abscissas of the set of matrices $\{J_{n_1,...,n_k} : (n_1,...,n_k) \geq (0,...,0)\}$ is positive and coincides with one of the eigenvalues of the matrix $J_{n_1,...,n_k}$, $\lambda^+_{n_1,...,n_k} = \Lambda > 0$.

Analogously, the reaction-diffusion equation (7) has an oscillatory instability near the fixed point $(\varphi_1, \varphi_2) = (0,0)$, if there exists some $(n_1,...,n_k) \geq (0,...,0)$ such that, $\text{Re}(\lambda^+_{n_1,...,n_k}) = \Lambda > 0$, with, $\text{Im}(\lambda^+_{n_1,...,n_k}) \neq 0$.

A necessary and sufficient condition for the existence of a Turing instability in a generic two-component system of reaction-diffusion equations has been obtained in [DIL 04b]. The connection between Turing instability and patterns will be discussed numerically in the next two subsections.

### B. *The Turing example*

In order to exhibit a system of reactions leading to stable patterns in a spatially extended media, Turing introduced the following set of hypothetical reactions:

$$
\begin{aligned}
& Y + X \xrightarrow{k_1} W \\
& W + A \xrightarrow{k_2} 2Y + B \quad (fast\ reaction) \\
& 2X \xrightarrow{k_3} W\ ;\ A \xrightarrow{k_4} X\ ;\ Y \xrightarrow{k_5} B \\
& Y + C \xrightarrow{k_6} C_s \quad (fast\ reaction) \\
& C_s \xrightarrow{k_7} X + C
\end{aligned}
\qquad (13)
$$

where $C$ is a catalyst. By the mass action law, to these reactions corresponds the set of evolution equations,

$$
\begin{aligned}
X' &= -k_1 XY - 2k_3 X^2 + k_4 A + k_7 C_s \\
Y' &= -k_1 XY + 2k_2 AW - k_5 Y - k_6 CY \\
W' &= k_1 XY - k_2 AW + k_3 X^2 \\
C' &= k_6 CY - k_7 C_s \\
C'_s &= -k_6 CY + k_7 C_s \\
A' &= -k_2 AW - k_4 A \\
B' &= k_5 Y + k_2 AW
\end{aligned}
\qquad (14)
$$

which have the conservation laws,

$$
\begin{aligned}
& A(t) + B(t) + C_s(t) + 2W(t) + X(t) + Y(t) = A(0) + B(0) + C_s(0) + 2W(0) + X(0) + Y(0) \\
& C(t) + C_s(t) = C(0) + C_s(0)
\end{aligned}
\qquad (15)
$$

Reactions rates (14) together with (15) represent a chemical process in a closed system. At this point, Turing assumed an open reaction by adding the condition $A = $ constant, and introduced the additional steady state conditions $\dot{W} = 0$ and $\dot{C} = 0$, which, from (14), leads to the steady state relations,

$$k_2 AW = k_1 XY + k_3 X^2$$
$$k_6 CY = k_7 C_s \qquad (16)$$

After substitution of (16) into (14), we obtain,

$$X' = -k_1 XY - 2k_3 X^2 + k_4 A + k_7 C_s$$
$$Y' = k_1 XY + 2k_3 X^2 - k_5 Y - k_7 C_s$$

and the conservation laws (15) play no role in the chemistry of this new system. With the choices, $k_1 = 25/16$, $k_3 = 7/64$, $k_4 = 10^{-3}/16$, $k_5 = 1/16$, $k_7 = 55 \times 10^{-3}/32$, $A = 10^3$ and $C_s = 10^3(1+\gamma)$, the above local system of equations leads to the reaction-diffusion equation,

$$\begin{cases} \dfrac{\partial X}{\partial t} = -\dfrac{50}{32} XY - \dfrac{7}{32} X^2 + \dfrac{57}{32} + \dfrac{55}{32}\gamma + D_X \Delta X \\ \dfrac{\partial Y}{\partial t} = \dfrac{50}{32} XY + \dfrac{7}{32} X^2 - \dfrac{2}{32} Y - \dfrac{55}{32} - \dfrac{55}{32}\gamma + D_Y \Delta Y \end{cases} \qquad (17)$$

where the parameter $\gamma$ depends on the initial concentration of the catalyst present in the reaction, $D_X$ and $D_Y$ are diffusion coefficients and $\Delta$ is the (one-dimensional) Laplace operator.

For $\gamma > -1024/285 \approx -2.659$, the local system associated to the reaction-diffusion equation (17) has two fixed points in phase space, with coordinates,

$$(x_0, y_0) = (-\tfrac{25}{7} - \sqrt{1024 + 385\gamma}, 1), \quad (x_1, y_1) = (-\tfrac{25}{7} + \sqrt{1024 + 385\gamma}, 1)$$

The fixed point $(x_0, y_0)$ is a saddle point (unstable) and, for $-2.659 \approx -1024/285 < \gamma < (471367 - 28685\sqrt{457})/721710 \approx -1.196$, $(x_1, y_1)$ is a stable node. If $-1.196 \approx (471367 - 28685\sqrt{457})/721710 < \gamma < 2416/4455 \approx 0.542$, $(x_1, y_1)$ is a stable focus. The local system associated to the reaction-diffusion equation (17) has a saddle-node bifurcation for $\gamma_{sd} = -1024/285 \approx -2.659$, and a supercritical Hopf bifurcation for $\gamma_H = 2416/4455 \approx 0.542$.

For $\gamma = 0$, $D_X = 0$ and $D_Y = 0$, the fixed point $(x_1, y_1)$ is a stable focus and has coordinates $(x_1, y_1) = (1, 1)$ but, due to the choice $A = $ constant, the two-dimensional local vector field associated to (17) is stiff. The numerical analysis of the solutions of (17), with periodic boundary conditions, $D_X = 1/2$ and $D_Y = 1/4$, lead Turing to conclude that there is in fact a Turing type instability, but the solution of the reaction-diffusion equation (17) diverges to infinity in finite time. Despite the existence of this catastrophic instability, Turing found a four lobe pattern for finite integration time.

The original kinetic system (14), together with the conservation laws (15), has solutions that exist for all $t \geq 0$, and the concentrations have always non-negative values. The stiffness in the above example has been introduced by the steady state simplification ($\dot{W} = 0$ and $\dot{C} = 0$), a technical artifact external to the original hypothetical mechanism of morphogenesis, and the qualitative analysis of the local system in (17) shows that local solutions, if they exist, can assume positive and negative values. However, Turing argued that the catastrophic instability could be eventually eliminated by non-linear effects or the exhaustion of the feeding components of the reaction ($A$ and $C$ for reaction (13)). In fact, it is difficult to understand how chemical processes in biological systems could be associated with dynamical processes having catastrophic instabilities.

*C. The Brusselator and the Ginzburg-Landau reaction-diffusion models*

To overcome the catastrophic instability associated to the Turing example, Prigogine and Lefevre [PRI 68], introduced the Brusselator model. According to these authors, this toy model mimics an autocatalytic processes and has the kinetic mechanism,

$$A \xrightarrow{k_1} X \ ; \ B + X \xrightarrow{k_2} Y + D$$
$$2X + Y \xrightarrow{k_3} 3X \ ; \ X \xrightarrow{k_4} E \tag{18}$$

where $X$ is the autocatalytic chemical substance. Applying the mass action law to (18), we obtain the system of differential equations,

$$X' = k_1 A - k_2 BX + k_3 X^2 Y - k_4 X$$
$$Y' = k_2 BX - k_3 X^2 Y \tag{19}$$
$$A' = -k_1 A \ ; \ B' = -k_2 BX \ ; \ E' = k_4 X \ ; \ D' = k_2 BX$$

with the conservation laws,

$$B(t) + D(t) = B(0) + D(0)$$
$$X(t) + Y(t) + A(t) + E(t) = X(0) + Y(0) + A(0) + E(0) \tag{20}$$

Assuming that $A$ and $B$ are constants, they are supplied during the reaction (open flow reactor experiments), the reaction-diffusion Brusselator model becomes,

$$\begin{cases} \dfrac{\partial X}{\partial t} = k_1 A - k_2 BX + k_3 X^2 Y - k_4 X + D_1 \Delta X \\ \dfrac{\partial Y}{\partial t} = k_2 BX - k_3 X^2 Y + D_2 \Delta Y \end{cases} \tag{21}$$

where $D_1$ and $D_2$ are diffusion coefficients, and $\Delta$ is the Laplace operator. The local component of the vector field associated to equation (21) has one fixed point with coordinates $(X_0, Y_0) = (Ak_1/k_4, Bk_2 k_4 / Ak_1 k_3)$. If $B < k_4/k_2 + A^2 k_1^2 k_3 / k_2 k_4^2$, $(X_0, Y_0)$ is a stable focus. If $B > k_4/k_2 + A^2 k_1^2 k_3 / k_2 k_4^2$, $(X_0, Y_0)$ is an unstable focus, and the local vector field has a supercritical Hopf bifurcation for $B = k_4/k_2 + A^2 k_1^2 k_3 / k_2 k_4^2$.

To analyze the properties of the solutions of equation (21), we first determine the regions on a parameter space where the Brusselator reaction-diffusion equation has Turing instabilities, [DIL 04b].

In Fig. 2, we show, in the parameter space, the regions of Turing instabilities and Turing patterns for the Brusselator model (21). In the presence of a Turing instability associated to the fixed point $(X_0, Y_0)$, we have searched numerically for the regions in the parameter space where Turing patterns exist. This is done by numerical integration of equation (21) with a benchmarked explicit numerical method, [DIL 98], where optimal convergence to the solution of the continuous system is achieved if, $\Delta t \max\{D_1, D_2\} / \Delta x^2 = 1/6$. Numerical integration for different values of $B$ and $D_1$, leads to the line separating the dark- and light-grey regions of Fig. 2. In the dark-grey region, a small perturbation of the steady state of the extended system evolves into a time independent and spatially non-homogeneous steady state solution of the reaction-diffusion equation (Turing pattern). In the light-grey region, the Brusselator model has a Turing instability but the asymptotic solutions are time periodic. On both sides of the Hopf bifurcation, the Brusselator model has Turing pattern solutions.

In the bifurcation diagram of Fig. 2, Turing patterns appear if $D_1 < D_2$. By (21), near the steady state $(X_0, Y_0)$, $Y$ is the activator and $X$ is the inhibitor, and the diffusion coefficient of the activator is larger than the diffusion coefficient of the inhibitor. If $D_1 > D_2$, we have Turing instabilities, but numerical integration of the Brusselator model (21) has not revealed Turing patterns. The diagram of Fig. 2 shows that, in the Brusselator model, the existence of a Turing instability is not sufficient to ensure the existence of Turing pattern solutions of reaction-diffusion equations. Moreover, when a Turing patterns exists, it is possible to change the initial conditions of the Brusselator model in such way that the asymptotic solutions of the reaction-diffusion equation is oscillatory in time. This shows that oscillatory and stable structures can coexist in the same model.

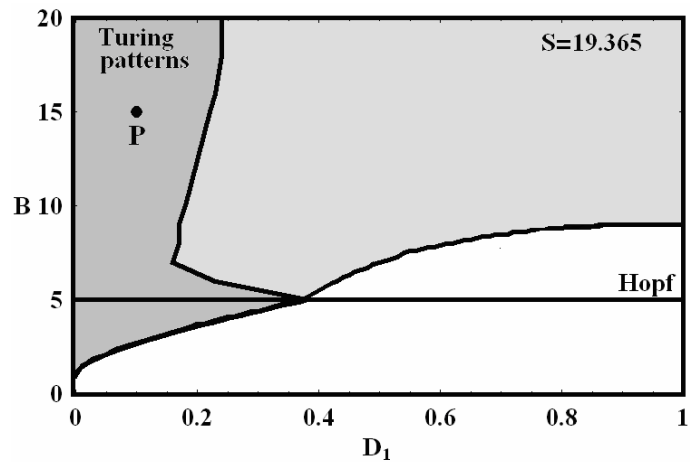

Fig. 2 Bifurcation diagram of the solutions of the Brusselator model (21), for the parameter values $A = 2$, $k_1 = k_2 = k_3 = k_4 = 1$ and $D_2 = 1$, in a one-dimensional domain of length $S = 19.365$, adapted from [DIL 04b]. The line $B = 5$ corresponds to the supercritical Hopf bifurcation of the fixed point $(X_0, Y_0)$ of the local system. Both grey regions correspond to parameter values where the origin has a Turing instability. Numerical integration of equations (21) for initial conditions deviating slightly from the steady state $(X_0, Y_0)$, shows that Turing patterns only appear in the dark-grey region. If $B > 5$, in the light-grey and white regions, asymptotic solutions are constant along the spatial region and oscillatory in time, with the period of the limit cycle. If $B \leq 5$, in the white regions, the asymptotic solutions are time independent and constant along the spatial region.

In Fig. 3, we show the asymptotic solution of the Brusselator model for the control point P of Fig. 2, in one- and two-dimensional spatial domains. Fourier analysis of the Turing pattern shows that the number of spatial periods in the pattern is not related with the most unstable real eigenmode. In this case, the eigenmode of order 10 has the largest eigenvalue, the eigenmodes from 0 to 35 are unstable with real eigenvalues, but the number of spatial periods in the Turing pattern is 7. Therefore, the number of spatial periods in the Turing patterns is not related with the eigenmode with the largest real eigenvalue (the spectral abscissa of the infinite system of equations (10)).

Further numerical experiments with the Brusselator model show that, for other parameter values, we can have the same type of qualitative behavior as the one found in some experiments. For example, in Fig. 4, we show back refraction phenomena generated with the Brusselator model (21), [SAI 98]. In this case, we have chosen parameter values of the local system where a small perturbations along a thin straight region in space leads to traveling wave fronts, and we have measured the propagation velocity (phase velocity). Changing parameters in such a way that the unstable steady state is unchanged, we have created a second region characterized by a different phase velocity. Then, following numerically the solutions of the Brusselator model on both regions, the wave fronts created in the first region refract at the interface between both media. As both media are characterized by different propagation velocities, we obtain the usual refraction phenomena of wave optics. However, if the incidence angle of the incoming wave hits the interface with an angle larger than the Brewster angle, we observe a back refraction phenomena, and the back refracted wave propagates and interfere with the incident wave. The back refraction wave fronts make an angle with the interface equal to the Brewster angle. In the media that is characterized by larger phase velocity, the interface and the direction of propagation are perpendicular. This phenomenon of back refraction is different from the optic phenomenon associated with the wave equation, and is a property of diffusion waves, [SAI 98].

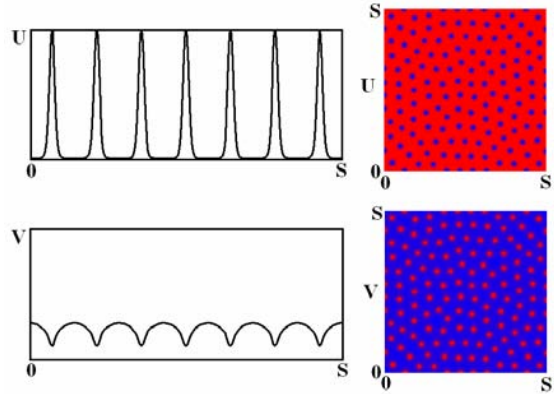

Fig. 3 Turing pattern solution of the Brusselator reaction-diffusion equation in a one- and a two-dimensional square of side length $S$, corresponding to the control point P of Fig. 2 ($D_1 = 0.1, B = 15$). The total integration time is $t = 1000$, with a time step of $\Delta t = 0.001$. We have chosen $N = 250$ lattice sites along each direction, implying that the length of the spatial domain is $S = N\sqrt{6\Delta t \max\{D_1, D_2\}} = N\Delta x = 19.365$. These solutions have been generated from a spatially extended random initial condition deviating slightly from the unstable fixed point of the local system. For $B = 15$, the local system has one unstable fixed point and one limit cycle in phase space. For other initial conditions, we obtain solutions that are asymptotically homogeneous in space and oscillatory in time. The period of the oscillations equals the period of the limit cycle of the local system. Adapted from [DIL 04b].

Back refraction phenomena and optical lens effects have been reported experimentally in the Belousov-Zhabotinsky reaction, [ZHA 93] and [PET 96]. The comparison between the numerical simulations with the Brusselator model and the experimental results show that wave type phenomena found in the Belousov-Zhabotinsky reaction have the signature of diffusion. The numerical analysis of these situations shows that back refraction and other optical phenomena only occur for parameter values where the unstable steady state is the same in both media, and the shape of the limit cycle in phase space do not change too much by variations of the parameters. However, the conditions on the steady state are difficult to check in real experiments.

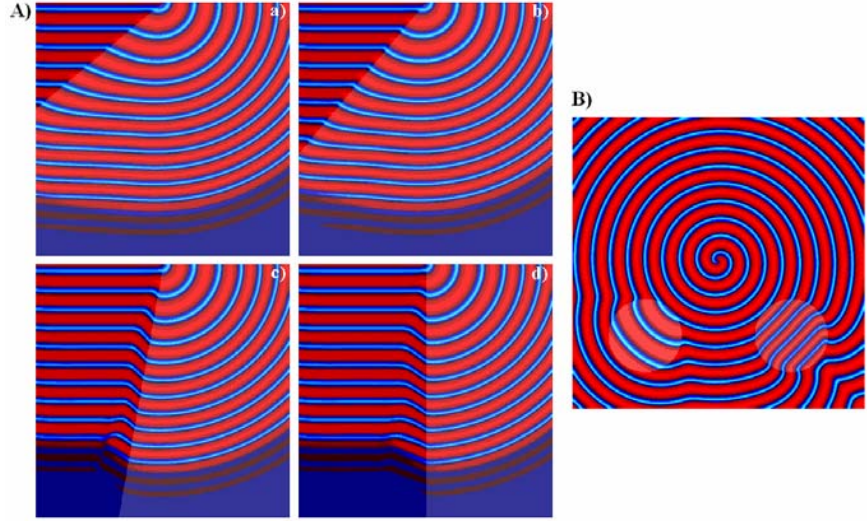

Fig. 4 Refraction phenomena calculated with the Brusselator model (21). A) The media on the left has parameter values: $A = 2$, $B = 4.6$, $k_1 = k_2 = 0.5$ $k_3 = k_4 = 1$, $D_1 = 1$ and $D_2 = 0$. In the media on the right we have changed some of the parameters to the new values: $k_2 = 0.54$ and $k_3 = 1.08$. The unstable steady state is the same in both media. The propagation speeds in both media are $c_{left} = 1.23$ and $c_{right} = 1.32$. In a) and b) the incidence angles are $\phi = 40°$ and $\phi = 50°$. The critical angle calculated by the Snell law is $\beta = 59.2°$. In c) and d), the incidence angles are $\phi = 80°$ and $\phi = 90°$, and the back refraction phenomena appears. In B), we show the optical lens effect simulated with the Brusselator model. In the media, we have the parameter values: $k_2 = 0.54$ and $k_3 = 1.08$. In the left spot, $k_2 = 0.58$ and $k_3 = 1.16$. In the right spot, $k_2 = 0.5$ and $k_3 = 1.0$. For details see [SAI 98]. Compare these simulations with the experimental results presented in [ZHA 93] and [PET 96].

As the Brusselator model and the Turing example of subsection B have in common a Hopf bifurcation, pattern formation properties of two-component systems of reaction-diffusion equations are better studied through the normal form or versal unfolding of the Hopf bifurcation of the local system. This reaction-diffusion equation has the form,

$$\begin{cases} \dfrac{\partial \varphi_1}{\partial t} = \nu \varphi_1 - \beta \varphi_2 + (\varphi_1^2 + \varphi_2^2)(a\varphi_1 - b\varphi_2) + D_1 \dfrac{\partial^2 \varphi_1}{\partial x^2} \\ \dfrac{\partial \varphi_2}{\partial t} = \beta \varphi_1 + \nu \varphi_2 + (\varphi_1^2 + \varphi_2^2)(a\varphi_2 + b\varphi_1) + D_2 \dfrac{\partial^2 \varphi_2}{\partial x^2} \end{cases} \quad (22)$$

System (22) has a supercritical Hopf bifurcation for $\nu = 0$, and a stable limit cycle exists for $\nu > 0$.

One of the features of the system of reaction-diffusion equations (22), also known as the Ginzburg-Landau reaction-diffusion equation, is its rotation invariance property. For this system, any initial condition with a rotationally symmetric initial condition evolves to a spiral, similar to the spirals shown in Fig. 1. For example, in Fig. 5, we show a spiral pattern obtained from (22) similar to the one- and two-arms spirals obtained experimentally in the Belousov-Zhabotinsky reaction, [AGL 94]. In Fig. 5d), we shown a pinwheel solution of (22), similar to the ones obtained experimentally by Lázár *et. all.*, [LAZ 95].

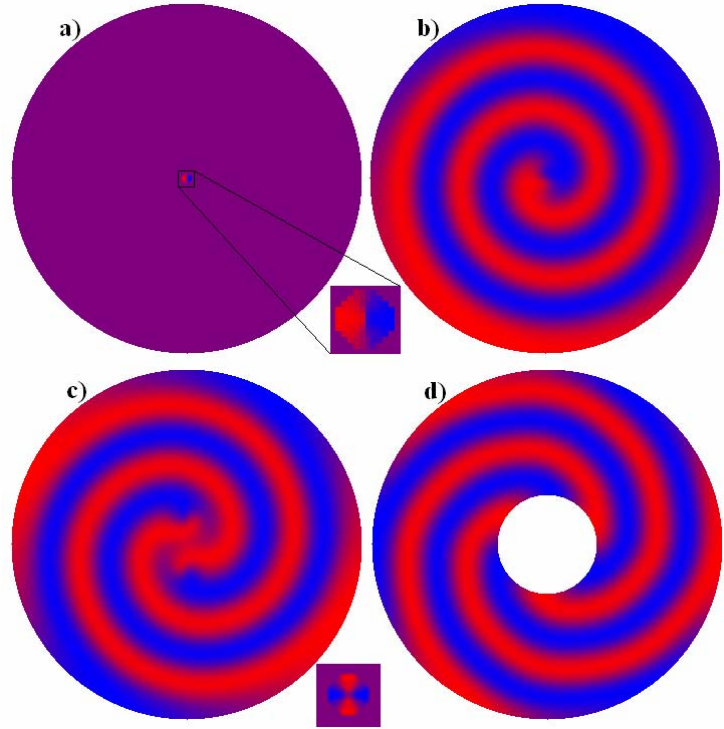

Fig. 5 Solutions $\varphi_1(x, y, t)$ of the reaction-diffusion equation (22) in a two-dimensional circular domain with zero flux boundary conditions. Blue represents higher values of $\varphi_1$ and red lower values. The circular region is inside a square lattice of 350 by 350 sites. The spatial scale is determined by the relation, $\Delta x = \sqrt{6D\Delta t}$, where $D = \max\{D_1, D_2\}$. Parameter values of the simulations are: $D_X = D_Y = 2\times 10^{-5}$, $\nu = 1$, $\beta = 1$, $b = 1$, $a = -1$, and $\Delta t = 0.01$. a) Initial conditions with $C_1$ symmetry around the central point. b) Spiral pattern developed from the initial condition in a). c) A two-arms spiral developed from an initial condition with $C_2$ symmetry. In the small square we show the enlargement of the central initial perturbation. d) A pinwheel rotating pattern generated by an initial condition with $C_3$ symmetry. In b)-c), the total integration time is $T = 30$, and in d), $T = 20$. For larger integration times, and after spiral arms hit the boundary, spirals become to unwind, and asymptotically the solutions of the reaction-diffusion system oscillates with the period of the limit cycle of the local vector field in (22). These patterns should be compared with the experimental results of [ZAI 70], [AGL 94] and [LAZ 95].

Both the Brusselator and the Ginzburg-Landau models show the most important patterning features as observed in the Belousov-Zhabotinsky reaction. However, there are some detailed aspects that are difficult to explain. For example, in the evolution sequence of a spiral of Fig. 1, the outer arm is fixed, whereas in the simulations of Fig. 5, the outer arm rotates with a constant angular velocity. Another feature of the Ginzburg-Landau reaction-diffusion equation is the non existence of Turing instabilities and patterns for parameter values when the unique fixed point of the local system is stable, [DIL 04b]. On the other hand, the Ginzburg-Landau equation (22) has Turing pattern solutions for parameter values where no Turing instabilities exist ([DIL 04b]), and the concept of Turing instability can not be taken as neither a necessary and sufficient condition for the existence of Turing patterns in systems of reaction-diffusion equations. On the other hand, other types of collective phenomena can be found in the Ginzburg-Landau equation as chemical turbulence and traveling isolated pulse patterns similar to the ones found in the Hodgkin-Huxley system describing pulse propagation along axons, [DIL 04c].

The comparison between the topologies of the solutions of the Ginzburg-Landau and the Brusselator reaction-diffusion equations suggests that the topology of patterns are very similar, but the classification advanced by Turing is not a criteria to decide on the different topologies of solutions of reaction-diffusion equations. On the other hand, the kinetic or local components of the Ginzburg-Landau and the Brusselator reaction-diffusion equations are topologically equivalent near the Hopf bifurcation, but the structure of the solutions of the associated parabolic equations do not share the same topological properties. This shows that the properties of the extended system cannot be derived from the properties of the local system.

This type of questions can only have a definitive answer from the precise calibration of the observed patterns in the Belousov-Zhabotinsky experiment with models.

### D. *Experimental facts about morphogenesis*

The first experimental fact that supports the Turing hypothesis has been reported by Waddington, [WAD 40]. In experiments on the development of embryos of chicken, dated from 1933, Waddington has shown that chemical messengers from certain tissues induce other tissues to develop.

Later, Lewis, Nüsslein-Volhard and Wieschaus, [NUS 96], have shown in a sequence of experiments that RNA of maternal origin diffuse in the embryo of *Drosophila*, establishing a gradient of concentration of bicoid RNA. In a sequence of mutation experiments, it has been shown that these gradients are the precursors of the segments in *Drosophila*.

However, experimental observations show that, in the early developmental stage (two hours after fertilization), the patterning found in the embryo of *Drosophila* is due to proteins that are localised around the nuclei in the syncytium, and have no mobility.

In the Belousov-Zhabotinsky experiment, patterning is observed due to the color change of the ferroin indicator which is a complex bounded to $Fe_3^{2+}$ (red) or $Fe_3^{3+}$ (blue), with a very low mobility in solutions. The experiments and simulations of optical phenomena in the Belousov-Zhabotinsky reaction show that reaction-diffusion has a role on the microscopic mechanisms of the reaction.

From the modeling point of view, the Brusselator model is constructed from a tri-molecular chemical mechanism, which is not very realistic. On the other hand, the Ginzburg-Landau equation is a pure formal reaction-diffusion equation and is not directly correlated with any microscopic mechanism. Both models show the same type of patterning as found in the Belousov-Zhabotinsky and in the social aggregates of the amoeba *Dictyostelium discoideum*, Fig. 6.

These facts suggest that reaction-diffusion equations are characterized by a finite number of patterns, and this can be seen as an universal property of this type of partial differential equations. On the other hand, diffusion and reaction-diffusion play an important role in the establishment of form and shape in the development of organism. With both hypothesis, and independently of the experimental calibration and validation of the models, we can understand why the same type of patterns appear in living systems in the class of reaction-diffusion equations.

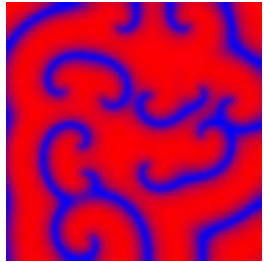

Fig. 6 Dynamic pattern obtained with the Brusselator reaction-diffusion equation (21) with Neumann boundary conditions and random initial data. Parameter values are: $A = 1$, $B = 2.2$, $k_1 = k_2 = k_3 = k_4 = 1$, $D_1 = 1$ and $D_2 = 0.1$. This pattern is similar to the ones found in the social aggregates of the amoeba *Dictyostelium discoideum.* Compare with [Gil 97].

III. BUTTERFLY EYESPOT FORMATION

Eyespots are concentric rings with contrasting colours on butterfly wings. They mimic the global appearance of vertebrate eyes and have active signalling functions against predators.

Lepidopterian wings are subdivided into cellular territories delimited by veins. Epithelial cells of butterfly wings form two juxtaposed cellular monolayers, corresponding to the dorsal and ventral wing faces. Wing surfaces are covered by mosaics of overlapping monochromatic scales spatially distributed as tiles on a roof. Scales are microscopic protuberances of specialized epithelial cells that display homogenous pigmentary coloration.

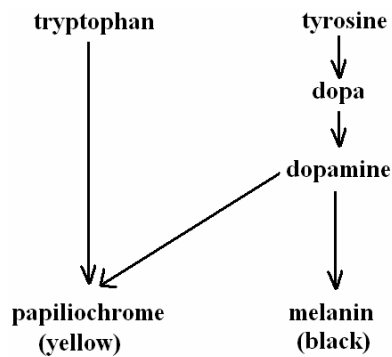

Fig. 7 Pigment precursors and their metabolic pathways. Adapted from [KOC 98].

The pigment coloration of butterfly wings results from selection of mutually exclusive metabolic pathways that synthesize only a particular type of pigment, [NIJ 91]. Studies with melanins and ommatins precursors (tyrosine and tryptophan, respectively), have shown that the tyrosine-melanin and the tryptophan-ommochromes pathways occurs in different areas of presumptive patterns. Tyrosine is incorporated preferentially in dark areas (black and dark-brown), whereas tryptophan is preferentially correlated with lighter colorations, [KOC 90], [KOC 91] and [KOC 98], Fig. 7.

Eyespot development involves different levels of biological organization, from molecular mechanisms to large scale evolution processes. It includes mechanisms of positional information, interaction between genetic and environmental factors, ecological adaptability and evolution. Selection studies have shown that eyespot patterns reveals high heritability, showing that phenotypic plasticity is under genetic control. On the other hand, eyespot development is affected by environmental factors such as temperature, humidity, photoperiod and substrate wing color, [NIJ 91].

Eyespot morphogenesis starts at larval stage, where the induction action of focus cells is corroborated by surgical experiments involving transplantation and destruction of focus cells in the early pupa stage, [BRA 96]. Local injury of focal areas produces reduction or total inhibition of eyespot formation, showing a close correlation between focus activity and eyespot development. Crafting of focus cells induces eyespot development in non-programmed areas of the wing. Both situations show the role of the eyespot focus as a source of a morphogenetic signal. We assume that eyespot development is triggered by localized messenger molecules that diffuse (morphogens) to nearby cells and react with other immobile substances, [DIL 04a]. The color motifs do not involve the messengers themselves but are generated by new substances formed in these chemical reactions, [DIL 04a].

Under these assumptions and based on the action of two diffusive substances or morphogens, we have proposed a reaction-diffusion mechanism for the formation of eyespots, [DIL 04a]. The first morphogen $M_1$ is produced in a specific region of the wing by an initial precursor $A$. The spatial distribution of the initial precursor $A = A(x, y, t = 0)$ at time zero defines the localisation of foci. The differentiation of the first ring (dark region) and the production of the second morphogen $M_2$ are induced by $M_1$ reacting with a wing background pigment precursor $P_0$ and modifying its chromatic properties. The second morphogen $M_2$ reacts with the wing background pigment precursor $P_0$, generating a second ring or light aureole. The final pigmentation is always defined by the highest local concentration of the pigment precursors.

This model can be implemented with the following kinetic pathways: 1) Focal cells release a primary diffusive morphogen $M_1$. The simplest conceivable kinetic mechanism is the following: $A \xrightarrow{k_1} M_1$, $M_1 \xrightarrow{k_2}$, where $k_1$ and $k_2$ are reaction rates, $A$ is the morphogen precursor, and the second reaction represents morphogen degradation. 2) The morphogen $M_1$ reacts with the background pigment precursor $P_0$ in the surrounding wing area, producing a new pigment precursor $P_1$ and a secondary morphogen $M_2$: $M_1 + P_0 \xrightarrow{k_3} P_1 + M_2$, $M_2 \xrightarrow{k_4}$. The pigment precursor $P_0$ is responsible for wing background pigmentation and $P_1$ for pigmentation of the first ring. 3) The diffusive morphogen $M_2$ produces a chemical modification in the pigmentation of the wing background: $M_2 + P_0 \xrightarrow{k_5} P_2$, where $P_2$ is the pigment precursor of the second ring. In Fig. 8, we schematically show this developmental mechanism.

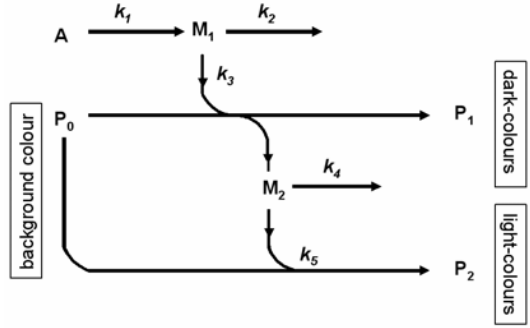

Fig. 8 Schematic diagram showing the several kinetic mechanisms considered in our model for the formation of butterfly wing eyespot patterns. Morphogens are diffusive substances and are represented by $M_1$ and $M_2$. Pigment precursors are $P_0$, $P_1$ and $P_2$. This mechanism mimics the metabolic pathways schematized in Fig. 7. Reprinted from [DIL 04a].

At the beginning of the developmental process, the primary pigment determinant $P_0$ is spatially distributed along the wing surfaces, except at focal cells. The position of the foci in the butterfly wing is described by the spatial distribution of the morphogen precursor $A$: focal areas are characterised by positive values of $A$, whereas outside focal areas $A = 0$. All other pigment determinants $P_1$ and $P_2$ are absent at the beginning of the developmental process.

Applying the mass action law to the model mechanisms of Fig. 8, we obtain the following system of (partial) differential equations:

$$\frac{\partial M_1}{\partial t} = k_1 A - k_2 M_1 - k_3 M_1 P_0 + D_1 \Delta M_1$$

$$\frac{\partial M_2}{\partial t} = k_3 M_1 P_0 - k_4 M_2 - k_5 M_2 P_0 + D_2 \Delta M_2 \qquad (23)$$

$$\frac{\partial A}{\partial t} = -k_1 A; \quad \frac{\partial P_0}{\partial t} = -k_3 M_1 P_0 - k_5 M_2 P_0; \quad \frac{\partial P_1}{\partial t} = k_3 M_1 P_0; \quad \frac{\partial P_2}{\partial t} = k_5 M_2 P_0$$

where $D_1$ and $D_2$ are the diffusion coefficient of morphogens $M_1$ and $M_2$, the $k_i$'s are reaction rate constants, and $\Delta = (\partial^2/\partial x^2 + \partial^2/\partial y^2)$ is the two-dimensional Laplace operator.

The system of equations (23) has the conservation law,

$$P_0(x, y, t) + P_1(x, y, t) + P_2(x, y, t) = \text{constant} \qquad (24)$$

and the spatial dependence of the pigment precursors is induced by the diffusion of the morphogens. In the steady state ($t \to \infty$), the concentrations of morphogens $M_1$ and $M_2$ vanish, and stable patterns emerge, obeying the conservation law,

$$P_0(x, y, t = 0) = P_0(x, y, t = \infty) + P_1(x, y, t = \infty) + P_2(x, y, t = \infty)$$

The spatial distribution of the pigment precursors at the steady state may be considered a pre-pattern that determines subsequent patterning events. This model presumes a competitive selection of exclusive pigment synthesis pathways within scale cells.

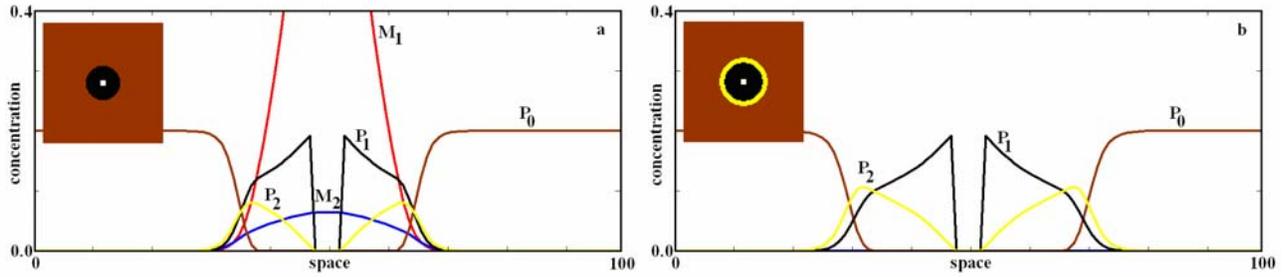

Fig. 9 Solution of the reaction-diffusion system of equations (23) in a square of $101 \times 101$ cells for time $t = 15$ (**a**), and the steady state ($t \to \infty$) (**b**). The graphs represent the concentration profiles of the model variables taken along a cross-section passing through the focus. All variables are represented in dimensionless form. The squares on the top left indicate the concentration levels of pigment precursors. The highest $P_0$ concentration areas are represented in brown. Black and yellow represent the highest concentrations of $P_1$ and $P_2$, respectively. The white spots represent focal points. Patterns emerge as the steady state solutions of equations (23) for the pigment precursors. At the steady state, the morphogen concentrations are zero. Simulations were carried out with zero flux boundary conditions and the following initial conditions: $P_1(x,y,0) = P_2(x,y,0) = 0$ and $M_1(x,y,0) = M_2(x,y,0) = 0$. The focus region corresponds to a square of $5 \times 5$ cells and is characterized by the initial conditions $A(x,y,0) = 20$ and $P_0(x,y,0) = 0$. Outside the focal region, $A(x,y,0) = 0$ and $P_0(x,y,0) = 0.2$. We selected the following simulation parameters: $k_1 = 1.0$, $k_2 = 0.05$, $k_3 = 4.0$, $k_4 = 0.01$, $k_5 = 4.0$, $D_1 = D_2 = 1.0$, $\Delta t = 0.1$ and $\Delta x^2 = 6 D_1 \Delta t$. $P_2$. Reprinted from [DIL 04a].

In order to test the above model, we have integrated numerically equations (23) with an explicit finite-difference method, minimizing the global numerical error, [DIL 98]. The simulations were performed with no-flux boundary conditions in a two-dimensional spatial region with 101x101 cells. The simulations start with uniform concentrations of the primordial pigment precursor ($P_0 = 0.2$) along the diffusion space, except at the focus areas where $P_0 = 0.0$. All the other variables are zero initially in the entire spatial range. In Fig. 9, we depict the time evolution of the morphogens and pigment precursors. The local conservation law (24) is responsible for the stability and patterning properties of the model (23).

According to this model, the formation of light ring of eyespot patterns is a consequence of the differentiation of the dark area. During the growth of dark areas, the concentration of the secondary morphogen increases, promoting the formation of light rings. The relative weight of light areas increases with the growth of global eyespot size. In Fig. 10, we show the dependence of eyespot dimensions as function of the focal area, as well as the patterns of interaction of eyespots. The interaction pattern in Fig. 10d, where the yellow aureole disappears between two close eyespot focus, is a common pattern that appears in some butterfly species, Fig. 11, and it is a typical signature of the reaction-diffusion mechanism of eyespots formation.

Eyespot variability may be understood within this framework. For example, by changing the position and width of a focus, it is possible to simulate different structures for eyespot patterns, as in Fig. 10d and 10e. Seasonal polyphenism and phenotypic plasticity are simulated by changing the rate constants of the model equations. In nature, this adaptability depends on environmental factors, such as temperature, relative humidity, and photoperiod, changing the reaction rates of kinetic mechanisms. Additional concentric rings may be modelled by the inclusion of new focus regions with different geometries, Fig. 11.

One of the novelties of this model is based on the assumption that the reaction-diffusion mechanism associated with the morphogens is the process that triggers the synthesis of the non-diffusive pigment precursor responsible for patterning. In the steady state, the concentration of morphogens is zero. This contrasts with the classical Turing reaction-diffusion approach, where patterning is due to spatial distribution of the morphogens at the steady state ([NIJ 91], [TUR 52] and [MUR 89]). Moreover, the results presented are consistent with the chemistry of all the major pigments in butterflies, Fig. 7.

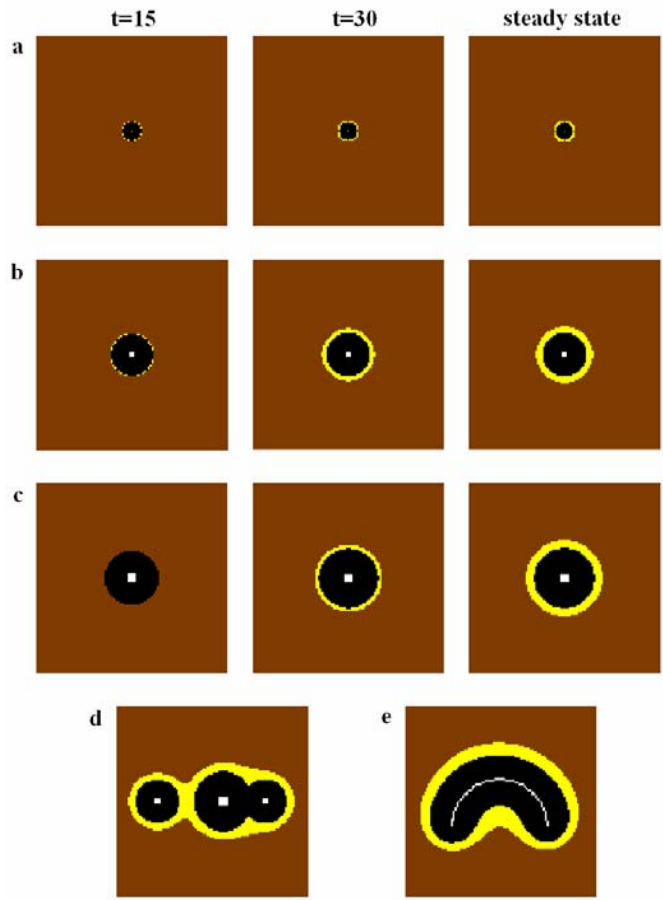

Fig. 10 From **a** to **c** we depict the time evolution of eyespot patterns as a function of the focal area, for the same parameters values of Fig. 9. In a), the focus region corresponds to one cell; in b), to a square of $3\times 3$ cells, and in c) to a square of $5\times 5$ cells. According to experimental observations, the larger the focus area, the larger the eyespot. In d) and e), we show that the variability and typical interaction patterns of eyespots for point (**d**) and arc (**e**) foci. The patterns in d) and e) are steady state solutions of equations (23). Reprinted from [DIL 04a].

These results can be interpreted as a memory effect induced by the conservation law (24). It can be easily shown that the system of equations (23) can be partially integrated and written in the form,

$$\frac{\partial M_1}{\partial t} = k_1 A(x,y,0)e^{-k_1 t} - k_2 M_1 - k_3 M_1 P_0 + D_1 \Delta M_1$$

$$\frac{\partial M_2}{\partial t} = k_3 M_1 P_0 - k_4 M_2 - k_5 M_2 P_0 + D_2 \Delta M_2$$

$$P_0(x,y,t) = P_0(x,y,0)\exp\left(-k_3 \int_0^t M_1(x,y,s)ds - k_5 \int_0^t M_2(x,y,s)ds\right) \quad (25)$$

$$P_1(x,y,t) = P_1(x,y,0) + k_3 \int_0^t M_1(x,y,s) P_0(x,y,s) ds$$

$$P_2(x,y,t) = P_2(x,y,0) + k_5 \int_0^t M_2(x,y,s) P_0(x,y,s) ds$$

The first three equations in (25) are independent from the two last solutions. Due to the integral terms in the third equation in (25), the asymptotic solution of P0 depends on the evolutionary history of the system of equations, which can be seen as a cumulative memory effect. This explains why, different initial conditions can give different topologically similar patterns. In the case of the Turing patterns generated by the Brusselator (and the Ginzburg-Landau equation, [DIL 04b]), small variations on the initial conditions do not change the final pattern. This explains the plasticity of this model in producing patterns that are robust and topologically similar.

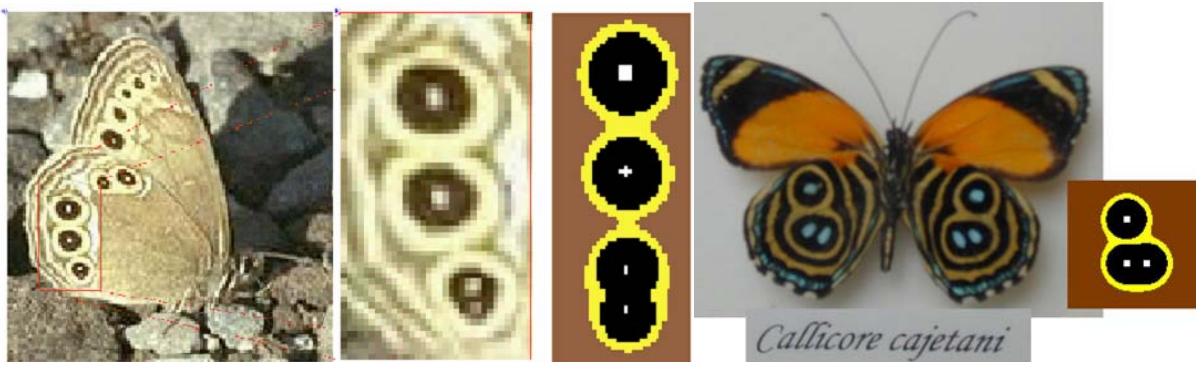

Fig. 11 Patterns in butterflies compared with patterns generated from equation (23).

IV. CONCLUSION

We have shown that reaction-diffusion equations, together with the mass conservation, generate a diversity of patterns that are very similar to the ones found in some biological systems, as butterfly eyespot patterns. Mass conservation introduces a memory effect in biological development, and phenotypic plasticity in patterns of living organisms can be explained by differences on the initial conditions occurring during development. This contrast with the usual Turing approach, where Turing patterns are invariant to small changes on the initial conditions, and mass conservation laws are clearly avoided after establishment of the local kinetic mechanisms.

From the observational and experimental point of view, patterning in biological systems are observed in the distribution of non diffusive substances, as is corroborated in the experiments with *Drosophila*, butterfly eyespots, *Dictyostelium Discoideum*, and the embryo of the chick. Models directly derived from the Turing original approach are derived from diffusive substances and are difficult to calibrate with observations and experiments. On the other hand, the similarity between the patterns obtained with the Brusselator and the Ginzburg-Landau reaction-diffusion equations with the patterns observed in the Belousov-Zhabotinsky reaction and in the amoeba *Dictyostelium Discoideum*, suggest that patterning in non-linear systems is a phenomena that is common to a large class of reaction-diffusion equations, independently of their connection with real systems.

ACKNOWLEDGMENTS

I would like to thank Rubem Mondaini for his kind invitation to deliver a talk at BIOMAT IV. This work has been partially supported by the POCTI program under the framework of the "Financiamento Plurianual das Unidades de I&D" from "Fundação para a Ciência e a Tecnologia" , Portugal.